\documentclass[prb,
twocolumn,
showpacs,
preprintnumbers,
amssymb, amsmath, floatfix]{revtex4}
\usepackage{graphicx}

\newcommand{\nn}{\ensuremath{\nonumber}}

\renewcommand{\vec}[1]{\ensuremath{\boldsymbol{\mathrm{#1}}}}
\newcommand{\grad}{\ensuremath{\nabla}}
\renewcommand{\div}{\ensuremath{\grad\cdot}}

\newcommand{\pd}[2]{\ensuremath{\frac{\partial #1}{\partial #2}}}

\usepackage{microtype}

\begin{document}
\title{Chiral anomaly and optical absorption in Weyl semimetals}
\author{Phillip E. C. Ashby}
\email{ashbype@mcmaster.ca}
\affiliation{Department of Physics and Astronomy, McMaster University, Hamilton, Ontario, Canada L8S 4M1}

\author{J. P. Carbotte}
\email{carbotte@mcmaster.ca}
\affiliation{Department of Physics and Astronomy, McMaster University, Hamilton, Ontario, Canada L8S 4M1}
\affiliation{The Canadian Institute for Advanced Research, Toronto, Ontario, Canada M5G 1Z8}

\begin{abstract}
Weyl semimetals are a three-dimensional topological phase of matter with isolated band touchings in the Brillouin Zone. These points have an associated chirality, and many of the proposals to detect the Weyl semimetal state rely on the chiral anomaly. A consequence of the chiral anomaly is that under the application of an $\vec{E}\cdot\vec{B}$ field, charge is transferred between points of opposite chirality.  In this paper we propose an optical absorption experiment that provides evidence for the chiral anomaly.  We use the Kubo formula, and find that an applied $\vec{E}\cdot\vec{B}$ induces the formation of step-like features at finite frequency in the interband optical conductivity.  We study the effect of scattering, and finite temperatures on this feature and find that it should be observable at low temperatures in pure samples.  Finally we discuss how the application of an $\vec{E}\cdot\vec{B}$ field can be used to map out the frequency dependence of the scattering rate. 
\end{abstract}

\maketitle
\section{Introduction}

Weyl semimetals are a novel topological phase of matter and have attracted considerable interest.\cite{Wan:2011fk,Hosur:2012fk,Witczak-Krempa:2012ve,vafek:2013fk} A Weyl semimetal is a three-dimensional system whose band structure contains pairs of bands crossings (called Weyl points) at isolated points in the Brillouin zone (BZ). For such a band crossing to occur, the Weyl semimetal state must break time-reversal or inversion symmetry. Each Weyl point can be assigned a chirality, $\chi$, that takes values $\pm1$. The Nielsen-Ninomiya theorem\cite{Nielsen:1981fv} shows that the number of Weyl points in the BZ must be even, with half the points of each chirality. Near a Weyl point with chirality $\chi$, the Hamiltonian takes the form
\begin{align}
\label{eq:hamil}  \mathcal{H} = \chi v_F \vec{k}\cdot\vec{\sigma},
\end{align}
where $\vec{k}$ is the momentum measured from the Weyl point, and $\vec{\sigma}$ is the vector of Pauli matrices. The three-dimensional nature ensures that the Weyl points are stable against perturbations. In fact, the only way to annihilate a Weyl point is if two points of opposite chirality meet in the BZ.

There are many candidate materials for the Weyl semimetal, yet there is still no compelling experimental evidence for the observation of one.  The pyrochore iridates\cite{Wan:2011fk,Witczak-Krempa:2012ve}, as well as topological insulator heterostructes\cite{Burkov:2011kx,Burkov:2011ys,Zyuzin:2012zr,Halasz:2012ly} were among the first systems proposed to host the Weyl semimetal state. It is also possible that certain quasicrystals may be host to the Weyl semimetal state\cite{Timusk:2013fk}.  The observation of linear conductivity over a wide frequency range is a sign of Dirac physics.  When combined with the lack of inversion symmetry there are sufficient conditions for the existence of the Weyl semimetal state. Recently there has been evidence for three dimensional Dirac physics in both Cd$_3$As$_2$ \cite{Neupane:fk2013,Borisenko:fk2013}and Na$_3$Bi\cite{xu:2013nb}.  The discovery of three dimensional Dirac materials is a promising first step towards the discovery of a Weyl semimetal.

Ideally, to correctly identify the Weyl semimetal, one needs as many probes as possible that can uniquely identify it from other phases (such as three dimensional Dirac semimetals).  Most of the research to date has focused on anomalous properties that can be traced back to the chiral anomaly\cite{Zyuzin:2012fkk,Grushin:2012fk,Aji:2012fk,Heon:2013fk,Liu:2013fkk,Landsteiner:2014fk}. The chiral anomaly is a peculiar non-conservation of chiral charge and has been mostly discussed in the context of high-energy physics. The Weyl semimetal is a condensed matter realization of the chiral anomaly and adds to the growing list of high-energy phenomenon in condensed matter systems\cite{Volovik}.  In the presence of external fields $\vec{E}$ and $\vec{B}$ the continuity equation for a Weyl point of chirality $\chi$ takes the form
\begin{align}
 \label{eq:contin} \pd{n}{t}+\div{\vec{j}} = \frac{\chi}{4\pi^2}\vec{E}\cdot\vec{B}.
\end{align} 
That is, the charge density at a single Weyl point is not conserved in the presence of parallel $\vec{E}$ and $\vec{B}$ fields. The missing (extra) charge at a given Weyl point is compensated at another Weyl point of the opposite chirality to ensure the overall conservation of charge in the system. Thus the application of parallel $\vec{E}$ and $\vec{B}$ fields can be used to drive charge between Weyl points of opposite chirality.  This charge pumping would continue until it is cutoff by some relaxation time, $\tau$ that corresponds to scattering between the two Weyl points.  As these scattering processes generically involve large momenta, $\tau$ is expected to be large\cite{hosur:2014fk}.

Most of the proposals to detect a Weyl semimetal have focused around experiments hoping to detect the chiral anomaly in some form.  One of the transport properties tied to the chiral anomaly is a large longitudinal magnetoconductivity\cite{Gorbar:2014fk}.  Other transport predictions have focused on an anomalous non-quantized Hall effect\cite{Xu:2011kx,Goswami:2012fk,Burkov:2011kx,Zyuzin:2012zr}, that is proportional to the separation of the Weyl points in momentum space.  The chiral magnetic effect is another consequence of the chiral anomaly wherein current flows parallel to an applied magnetic field\cite{Jian-Hui:2013kx,Chen:2013uq}.  Another transport experiment was proposed in which chiral charge pumping could be measured as a voltage drop over long distances\cite{Parameswaran:fk2013}.  There has also been a proposal to measure the chiral anomaly through the non-vanishing gyrotropy induced by external fields\cite{hosur:2014fk,Goswami:2014fk}.  Most recently, the density response of Weyl semimetals was studied\cite{Pesin:2014fk}, showing that both the compressibility and plasmon modes contain signatures unique to the Weyl semimetal state.

In this paper, we propose an optical absorption experiment that measures the chiral anomaly. Our proposal takes advantage of the charge pumping induced by the chiral anomaly.  The charge pumping leaves different Weyl points at different chemical potentials and causes measurable effects in both the Drude peak, as well as in the interband portion of the conductivity.  In particular, we identify sharp step-like features in the interband conductivity that should prove as another `smoking gun' for the Weyl semimetal state.

\section{Single point conductivity}
\begin{figure}
\centering
  \includegraphics[width=0.8\linewidth]{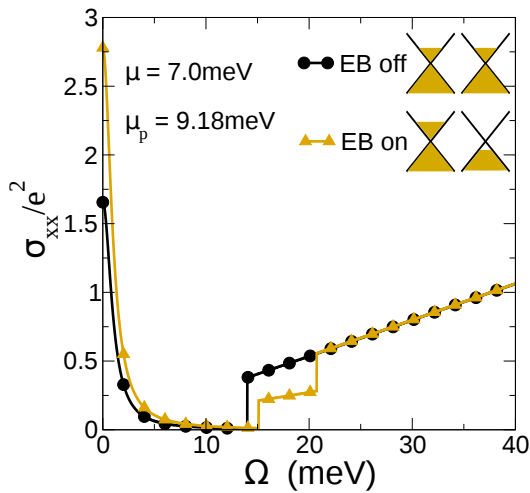}
    \caption{(Color online) The finite frequency optical conductivity for a clean Weyl semimetal at $T=0$. In black we show the optical conductivity for a doped Weyl semimetal, we have used a small broadening, $\gamma = 0.01$, to the Drude for graphing purposes. After the application of the applied $\vec{E}\cdot\vec{B}$ field charge is pumped from one Weyl point to the other, and step-like signatures appear in the interband portion of the optical conductivity. The missing interband spectral weight is transferred to the Drude peak.  The measurement of these interband features is tied to the chiral anomaly and would be a direct signature of a Weyl semimetal.}
  \label{fig:main}
\end{figure}

Our starting point is the Kubo formula for the optical conductivity.  Written in terms of the spectral functions, $A$, and for a chemical potential, $\mu$, the $xx$ component of the real part of the optical conductivity is given by
\begin{align}
\nn  \sigma_{xx}(\Omega)= \frac{e^2\pi}{\Omega}&\int d\omega [f(\omega-\mu) - f(\omega-\mu+\Omega)]\\
\label{eq:kubo}  &\times\int \frac{d^3k}{(2\pi)^3}\mathrm{Tr}\left[v_x\hat{A}(\vec{k},\omega)v_x\hat{A}(\vec{k},\omega+\Omega)\right].
\end{align}
Here $f(x) = 1/(e^{x/T}+1)$ is the Fermi function, and $v_x = \pd{\mathcal{H}}{k_x} = \sigma_x$ are the velocity operators. We work in units where $\hbar = v_F = k_B = 1$ and all photon energies and temperatures are in meV (in Appendix \ref{sec:apa} we restore the factors of $v_F$ and $\hbar$ for clarity).  The spectral functions are found from the decomposition of the Green's function
\begin{align}
  \hat{G}(\vec{k},\omega) = \int d\omega' \frac{\hat{A}(\vec{k},\omega')}{\omega-\omega'}.
\end{align}
After evaluating the trace and converting the $\vec{k}$-space integration to an integral over energy, the optical conductivity appears as the sum of two terms, $\sigma_{xx} = \sigma^\textrm{D}+\sigma^\textrm{IB}$. Details of this derivation are provided in Appendix \ref{sec:apa}.  The first term is a Drude (or intraband) term, $\sigma^\textrm{D}$, and the second is an interband term, $\sigma^{\textrm{IB}}$. We find
\begin{align}
\nn  \sigma^{\textrm{D}}_{xx}(\Omega) = \frac{e^2}{6\pi}\int_{-\infty}^\infty d\omega \frac{f(\omega-\mu)-f(\omega-\mu+\Omega)}{\Omega}\\\times\int_0^\infty d\epsilon \epsilon^2\left[A(\epsilon,\omega)A(\epsilon,\omega+\Omega)+A(-\epsilon,\omega)A(-\epsilon,\omega+\Omega)\right],
\end{align}
and
\begin{align}
\nn  \sigma^{\textrm{IB}}_{xx}(\Omega) = \frac{e^2}{3\pi}\int_{-\infty}^\infty d\omega \frac{f(\omega-\mu)-f(\omega-\mu+\Omega)}{\Omega}\\\times\int_0^\infty d\epsilon \epsilon^2\left[A(\epsilon,\omega)A(-\epsilon,\omega+\Omega)+A(-\epsilon,\omega)A(\epsilon,\omega+\Omega)\right].
\end{align}
We would like to point out the factor of two difference between the interband term and the Drude term. This factor of two arises when preforming the angular integration in Eq. (\ref{eq:kubo}).

In the presence of a self energy $\Sigma(\omega)$ the spectral functions are given by
\begin{align}
  A(\pm\epsilon,\omega) = \frac{1}{\pi}\frac{-\textrm{Im}\Sigma(\omega)}{\left(\omega-\textrm{Re}\Sigma(\omega)\mp\epsilon\right)^2+\left(\textrm{Im}\Sigma(\omega)\right)^2}.
\end{align}
We can now perform the integration over $\epsilon$ to obtain expressions for the conductivity. An essential feature of our expressions is that they retain the energy dependence of $\Gamma(\omega)$. We adopt the shorthand $\Gamma(\omega) = \Gamma$ and $\Gamma(\omega+\Omega) = \Gamma'$ for the frequency dependent scattering rate. In the limit of a small impurity scattering rate, and neglecting the real part of the self energy we obtain
\begin{align}
\nn\sigma^\textrm{D}(\Omega)=\frac{e^2}{6\pi^2}\int_{-\infty}^\infty d\omega\frac{f(\omega-\mu)-f(\omega+\Omega-\mu)}{\Omega}\\
\label{eq:drude}\times\frac{\omega^2\Gamma'+(\omega+\Omega)^2\Gamma}{(\Gamma+\Gamma')^2+\Omega^2}
\end{align}
for the Drude piece, and
\begin{align}
\nn\sigma^\textrm{IB}(\Omega)= \frac{e^2}{3\pi^2}\int_{-\infty}^\infty d\omega\frac{f(\omega-\mu)-f(\omega+\Omega-\mu)}{\Omega}\\
\label{eq:ib}\times\frac{\omega^2\Gamma'+(\omega+\Omega)^2\Gamma}{(\Gamma+\Gamma')^2+(2\omega+\Omega)^2}
\end{align}
for the interband piece.

In the strict $\Gamma = 0$ and $T = 0$ limit our formula reduces to the well known result
\begin{align}
\label{eq:clean}\sigma_{xx}(\Omega)= \frac{e^2\mu^2}{6\pi v_F}\delta(\Omega)+\frac{e^2\Omega}{24\pi v_F}\Theta(\Omega-2|\mu|),
\end{align}
where we have restored the factor of the Fermi velocity that defines the Weyl Fermions (see also Eq. (\ref{eq:a14}) and (\ref{eq:a16})).  This makes it clear how the relativistic Hamiltonian Eq. (\ref{eq:hamil}) impacts the conductivity.

For a moment let us consider only the intraband term, $\sigma^\textrm{D}$.  In the small $\Omega$ limit (appropriate for this term) we have $\Gamma = \Gamma'$ and for $\Omega \ll T$ the difference $[f(\omega-\mu)-f(\omega+\Omega-\mu)]/\Omega$ can be replaced by $-\pd{f}{\omega}$. In this limit we have 
\begin{align}
\sigma^\textrm{D}(\Omega) = \frac{e^2}{3\pi}\int_{-\infty}^\infty d\omega\left(-\pd{f}{\omega}\right)\omega^2\frac{\Gamma}{4\Gamma^2+\Omega^2},
\end{align}
which is precisely the form of the Boltzmann equation for the conductivity used elsewhere in the literature\cite{Burkov:2011ys}.

\subsection{Weyl semimetal}

Now that we have obtained formulae for a single Weyl point we can examine the consequences in the Weyl semimetal state.  A real Weyl semimetal contains an even number of Weyl points, with half of each chirality.  For our purposes it will be sufficient to consider the case of two Weyl points.  The chiral anomaly implies that the application of a constant $\vec{E}\cdot\vec{B}$ will induce a charge difference between the pair of Weyl points. This change in charge density is captured through a change in the chemical potential at each Weyl point.  The charge density continues to change until one reaches a steady state characterized by some relaxation time, $\tau$.  The pumping of the chemical potential due to the applied field is given by $\mu_p^3 = \frac{3e^2\hbar v_F^3}{2}\vec{E}\cdot\vec{B}\tau$ (we have restored the factors of $\hbar$ and $v_F$ in this equation for clarity). For a Weyl point of chirality $\chi$ and chemical potential $\mu$ before the application of the  $\vec{E}\cdot\vec{B}$ field, the resulting chemical potential after pumping is given by\cite{hosur:2014fk}
\begin{align}
\mu_\chi^3 = \left(\mu^3+\chi\mu_p^3\right),
\end{align}
and it is understood that the real root should always be chosen. To estimate the size of the chemical potential shift we take  $v_F = 4.3\times10^5$m/s, $B = 1$T, $E = 10^3$V/m and $\tau = 10^{-11}$s which gives $\mu_p =9.18$meV.  Such a chemical potential shift should be observable in a low frequency optical experiment. The value of $v_F$ stated above is typical of 2D Dirac systems\cite{Liu14:PRB,ZhouCarb:fk14} and is conservative for 3D Dirac materials. An experiment by Orlita {\it et al.}\cite{Orlita:2014fk} presents spectroscopic evidence for 3D Dirac Fermions with a velocity $v_F = 10^6$m/s. Timusk {\it et al.}\cite{Timusk:2013fk} recently pointed out that the quasicrystal AlCuFe and its related approximant Al$_2$Ru show a conductivity that is remarkably linear over a large energy range ($>$0.5eV).  Their best estimate of the Fermi velocity is $4.3\times10^6$m/s. Such large Fermi velocities ($v_F >10^6$m/s) leads to a chemical potential shift 10 times larger than our estimated energy.

In Figure \ref{fig:main} we show the finite frequency optical conductivity for two Weyl points in the clean limit ($\Gamma=0$) at $T=0$. In this case the optical conductivity is simply given by
\begin{align}
\label{eq:clean2}\sigma_{xx}(\Omega)= \frac{e^2}{6\pi}\sum_\chi\left[\mu_\chi^2\delta(\Omega)+\frac{\Omega}{4}\Theta(\Omega-2|\mu_\chi|)\right].
\end{align}
The Weyl points pictured have a chemical potential initially at $7$meV.  The optical conductivity is given by Eq. (\ref{eq:clean2}) (with $\mu_\pm =$ 7meV) and is shown in black.  After the application of the $\vec{E}\cdot\vec{B}$ field the chemical potentials change at the two Weyl points.  The resulting optical conductivity after charge pumping is pictured in orange.  As can be seen in Figure \ref{fig:main} the pumping has two effects: a step-like feature has appeared in the interband conductivity, and spectral weight has been transferred to the Drude peak.  Careful measurement of these two features as a function of applied $\vec{E}\cdot\vec{B}$ would be a direct signal of the Weyl semimetal state since this phenomenon is intimately linked with the chiral anomaly.

\begin{figure}
\centering
  \includegraphics[width=0.8\linewidth]{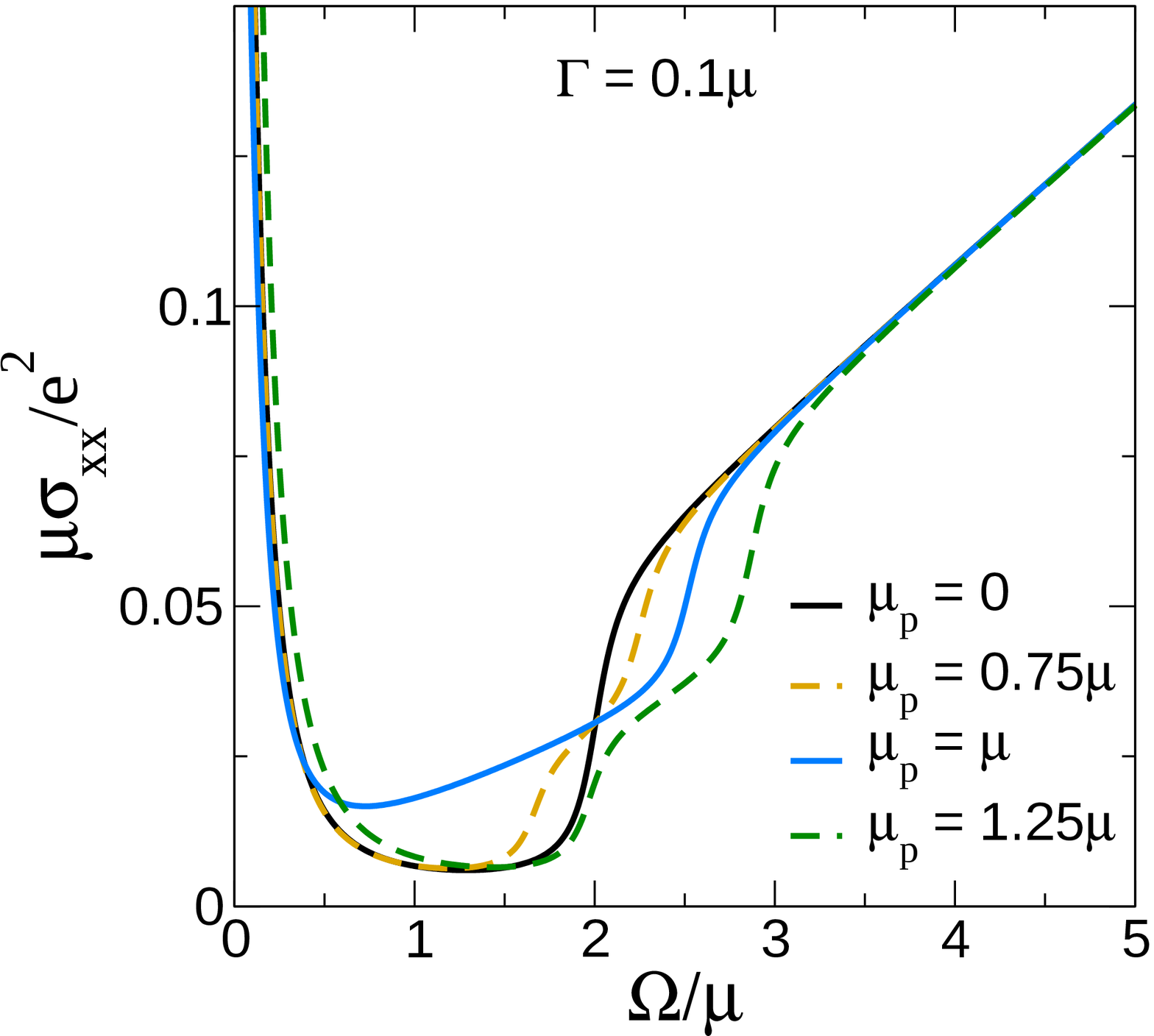}
  \includegraphics[width=0.8\linewidth]{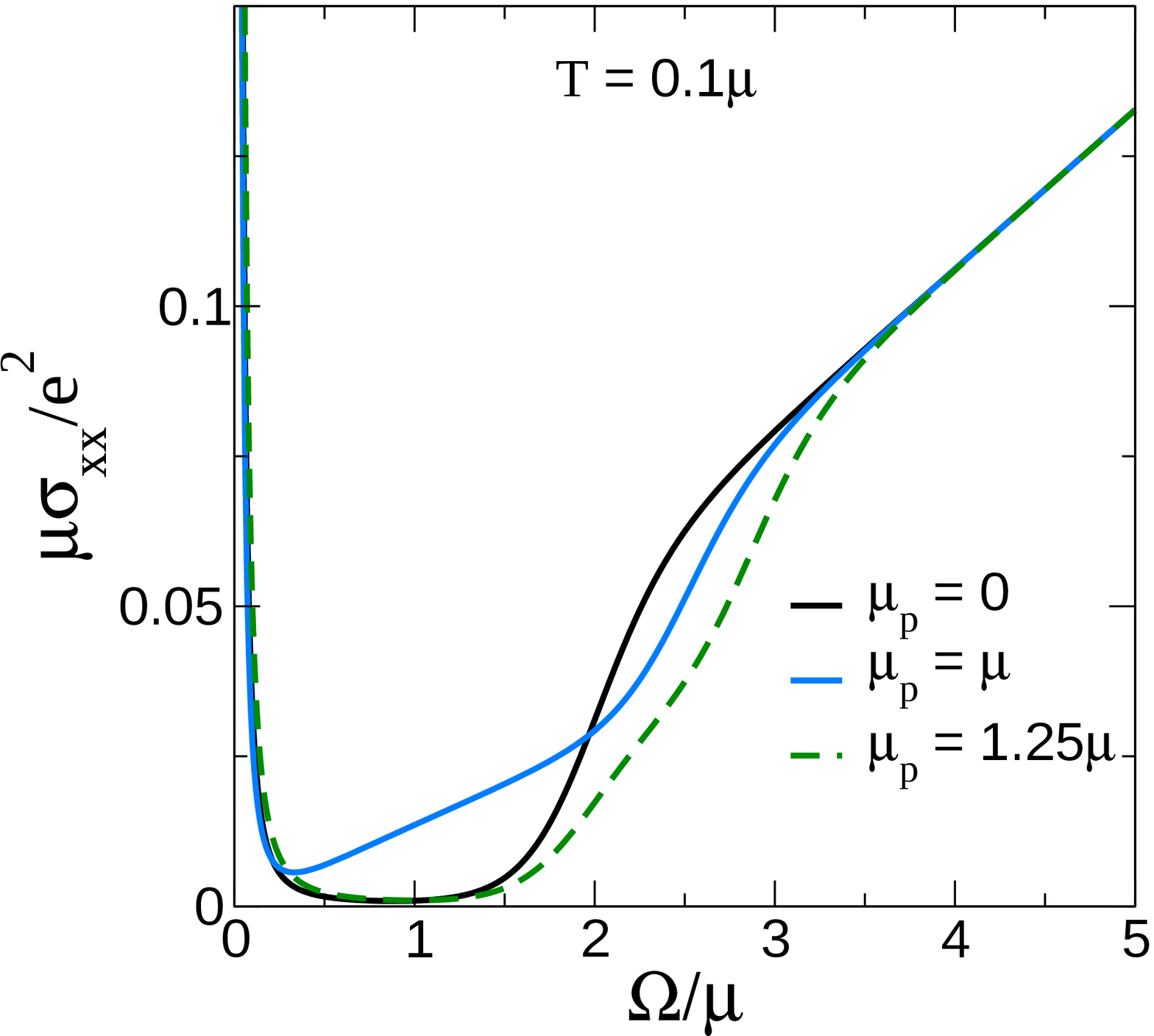}
    \caption{(Color online) Top: The finite frequency conductivity for several different values of the charge pumping. In this figure we show how disorder smears out the step-like features in the interband. The pictured curves are for a residual scattering rate of $\Gamma = 0.1\mu$. The step-like features are most prominent when $\mu_p \approx \mu$ and are clearly visible as long as the scattering is not an appreciable fraction of $\mu$. Also notice the anomalous case, $\mu_p = \mu$ which contains a linear piece all the way to zero frequency.
    Bottom:The finite frequency conductivity for several different values of the charge pumping at finite temperature. Finite temperature shifts the chemical potential downward, as well as introducing thermal broadening. This figure has $T=0.1\mu$, the same energy as we used in the residual scatting figure. However, the effect of temperature is much more noticeable, and the step-like features have almost been completely washed out.}
  \label{fig:effects}
\end{figure}

There are two special cases that we now mention. The first is an undoped Weyl semimetal.  In this case, the charge pumping simply transforms it into a doped Weyl semimetal with doping $\mu_p$ (one Weyl cone is electron doped, and the other is hole doped).  In this case there is a single step in the interband conductivity, and the missing spectral weight is transferred to the Drude.  The second case is when the chemical potential due to pumping, $\mu_p$ exactly matches the initial doping, $\mu$. In this case, one Weyl point is completely drained of its charge density.  The point that sits at charge neutrality has no Drude and contributes linearly at all frequencies. The resulting optical conductivity has a peculiar shape (see the blue curve in Fig \ref{fig:effects}). 

The presence of these steps can be understood from an inspection of Eq. (\ref{eq:clean}). Most simply, the steps are a consequence of the Pauli principle for the two absorption scales $\mu_\chi$. We would like to point out that Eq. (\ref{eq:clean}) was obtained in the analysis the Faraday and Kerr rotations presented by Hosur and Qi\cite{hosur:2014fk}. Our calculation of the Kubo bubble is essentially the same as theirs except that we have not taken the $T\rightarrow0$ or $\Gamma\rightarrow0$ limit. Indeed, the Faraday and Kerr effects can be derived from the optical conductivity.  Here we have chosen to focus on the direct measurement of the changes in the optical conductivity, rather than a derived quantity. The experiment outlined by Hosur and Qi\cite{hosur:2014fk} is both technically challenging, and produces an incredibly small signal (picoradians for the Kerr effect).  On the other hand, the chemical potential shift of 9meV occurs at frequencies routinely measured in infrared spectroscopic experiments\cite{RevModPhys.77.721}. Thus, we expect that a direct measurement of the chiral chemical potential shift through optical absorption will be favorable.

\subsection{Impurities}

We now turn to the effect of impurities on the features that we saw in Figure \ref{fig:main}.  It is important to understand if these features will still be observable in the presence of disorder. As a first approximation we take a constant residual scattering rate in Eqns. (\ref{eq:drude})  and (\ref{eq:ib}).  In this case, at $T=0$ we obtain simple expressions for the single node conductivity:
\begin{align}
\sigma^\textrm{D}(\Omega,T=0)=\frac{e^2\mu^2\Gamma}{3\pi^2(4\Gamma^2+\Omega^2)},
\end{align} 
and 
\begin{align}
\nn\sigma^\textrm{IB}(\Omega,T=0)=\frac{e^2}{24\pi^2}\left[4\Gamma+\Omega\textrm{arccot}\left(\frac{2\Gamma}{2\mu+\Omega}\right)\right.\\
\left.-\Omega\textrm{arccot}\left(\frac{2\Gamma}{2\mu-\Omega}\right)\right].
\end{align}
In the formula for $\sigma^\textrm{D}$ we have assumed $\Omega \ll \mu$. We see that in this case $\sigma^\textrm{D}$ takes the form of a simple Drude peak.  The conductivity is plotted as a function of frequency in the top panel of Figure \ref{fig:effects} for a constant residual scattering rate $\Gamma = 0.1\mu$ for several different values of the charge pumping. Notice the peculiar shape of the conductivity when $\mu_p = \mu$. At all values of $\mu_p$ the step-like feature at finite frequency has been smoothed out by disorder, but is still clearly visible. The step-like feature is most pronounced when $\mu_p \approx \mu$ due to the way the chemical potentials add, away from this region the larger of the two energies, $\textrm{max}[\mu,\mu_p]$, dominates the shape of the conductivity. 

Finally we considered the Born approximation, where the scattering rate is proportional to the density of states, $\Gamma(\omega) \propto g(\omega)$.  We found that the form of scattering had little effect on the interband features in the conductivity. The impact of the Born approximation on the Drude will be discussed later.

\begin{figure}
\centering
  \includegraphics[width=0.8\linewidth]{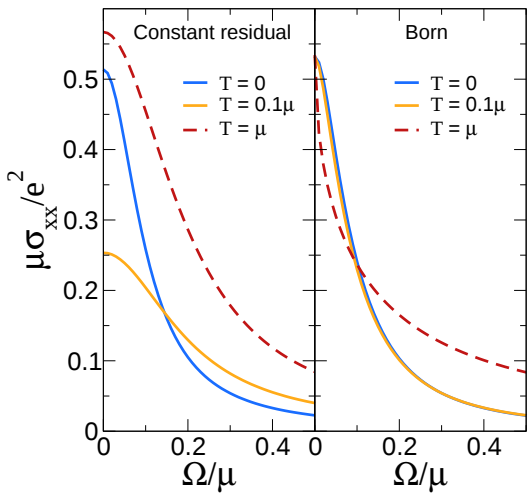}
    \caption{(Color online) Here we show the effect of finite temperature on the shape of the Drude peak for two different types of scattering.  Constant residual scattering is pictured on the left, and finite temperature simply causes the Drude peak to broaden. On the right we show the case of Born limit scattering.  In this case the scattering rate is frequency dependent and the lineshape depends on which energy is dominant. For $T<|\mu|$ the lineshape has a Drude form, while for $T > |\mu|$ the lineshape develops a sharp cusp.}
  \label{fig:drude}
\end{figure}

\subsection{Finite temperature}
Until now we have shown results at $T=0$.  A Weyl semi-metal has a non-constant density of states, which results in the chemical potential being strongly temperature dependent.  Since we are interested in the effects of finite temperature as well as finite doping, we include the shift in $\mu$ due to finite $T$.  To find the chemical potential as a function of temperature we require that the charge density remain constant as we change T.  The charge density is given by
\begin{align}
  n = \int \frac{d^3k}{(2\pi)^3}\left[f(\epsilon-\mu)-f(\epsilon+\mu)\right].
\end{align}

We use the identity
\begin{align}
\frac{df(\epsilon\pm\mu)}{dT} = \left(\frac{\epsilon\pm\mu}{T}\mp\frac{d\mu}{dT}\right)\left(-\pd{f}{\epsilon}\right)
\end{align}
and integrate over $k$.  Thus, we obtain the following differential equation for $\mu$:
\begin{align}
\frac{d\mu}{dT}\left(\mu^2+\frac{\pi^2T^2}{3}\right)+\frac{2\pi^2T}{3}\mu = 0.
\end{align}
Combining this with the boundary value $\mu(0) = \mu_0$ gives the following solution
\begin{align}
  \mu(T) = \frac{2^{1/3}\left(9\mu_0^3+\sqrt{81\mu_0^6+12\pi^6T^6}\right)^{2/3}-2\pi^23^{1/3}T^2}{6^{2/3}\left(9\mu_0^3+\sqrt{81\mu_0^6+12\pi^6T^6}\right)^{1/3}}.
\end{align}
This equation gives the chemical potential at finite $T$, which is rapidly suppressed as $T$ increases. Thus, finite $T$ has two effects on the conductivity: The first is the usual thermal broadening from the Fermi functions. The second is the finite $T$ shift, which moves the $\mu_p$ split chemical potentials closer to one another. Both of these effects tend to smear out the step-like features that identify the Weyl semimetal.

The conductivity at $T=0.1\mu$ is plotted as a function of frequency for several values of $\mu_p$ in the bottom panel of Figure \ref{fig:effects}.  The effect of temperature is much more drastic than disorder, even though the energy scale is similar.  Although the step-like feature is almost completely smeared out, the onset of spectral weight transfer from the interband to the Drude due to finite $\mu_p$ is still a clear signature of the Weyl semimetal state.  

Finite temperature also has a large effect on the shape of the Drude peak at $\Omega = 0$.  In Figure \ref{fig:drude} we show the Drude peak for both constant residual scattering (left) and Born limit scattering (right). For constant residual scattering, the peak always has a Drude from and the effect of finite temperature only broadens the peak further. In the Born limit the scattering rate is proportional to the density of states: $\Gamma(\omega) = \gamma\omega^2/\pi$, resulting in an unusual lineshape. For a constant scattering rate, the conductivity as a function of $\omega$ is concave down for small $\omega$ and approaches $\omega=0$ with zero slope.  This is in stark contrast to the dashed red curve in the right hand frame of Figure \ref{fig:drude} which shows a cusp at $\omega=0$ and is concave up. The unusual lineshape associated with Born limit scatterers was first pointed out by Burkov and Balents for the case of a $\mu = 0$ Weyl semimetal\cite{Burkov:2011kx}.  They showed that in this case the line shape had a cusp of the form
\begin{align}
\sigma^\textrm{D}(\Omega)  = \frac{e^2}{6\pi\gamma}\left(1-\frac{1}{8}\sqrt{\frac{2\pi^3\Omega}{\gamma T^2}}\right).
\end{align}
This lineshape follows from our expressions in the Born limit provided that $\Omega \ll T$. If the chemical potential is increased by the application of an $\vec{E}\cdot\vec{B}$ field so that $|\mu| > T$ then the lineshape for a single Weyl point takes the form
\begin{align}
\sigma^\textrm{D}(\Omega) = \frac{e^2\gamma}{3\pi^2}\left[\frac{\mu^4}{\frac{4\gamma^2}{\pi^2}\mu^4+\Omega^2}\right],
\end{align}
which has the form of a Drude peak. The width of the Drude peak can be increased simply by increasing the applied $\vec{E}\cdot\vec{B}$ field. Using the chiral anomaly to change $\mu_p$ allows one to change the energy scale at which $\Gamma(\omega)$ is probed.  In this way the frequency dependence of a general $\Gamma(\omega)$ can be completely mapped out, since $\Gamma(\omega)$ is responsible for the shape of $\sigma^\textrm{D}$. The spectral weight in the Drude peak characterizes its width. At finite $T$ and $\mu$ we find that the spectral weight is given by
\begin{align}
  \int_0^\infty d\omega\sigma(\omega) = \frac{1}{12\pi}\left(\mu^2+\frac{T^2\pi^2}{3}\right).
\end{align} 
The Drude weight does not vanish at $T=0$ since it is experimentally impossible to arrange $\mu=0$.

\section{Conclusions}

We have described a method for detecting the chiral anomaly using an optical absorption experiment. The chiral anomaly is one way in which the Weyl semimetal is distinct from its 3D-Dirac cousin. Direct measurement of the chiral anomaly is therefore a sign of bulk Weyl points. We show that there are signatures in the finite frequency optical response that can be controlled through the application of an applied $\vec{E}\cdot\vec{B}$ field.  After the application of an $\vec{E}\cdot\vec{B}$ field a pair of Weyl points have different chemical potentials and step-like features appear in the interband portion of the conductivity. An important result is that the signature of this anomaly in the absorptive part of the conductivity is large and should be easily detected with presently available optical absorption techniques. In addition to this, spectral weight is transferred to the Drude peak at $\Omega = 0$.  We showed that these features remain as long as the scattering rate remains small compared to the chemical potential. We estimated that the pumping provided by the $\vec{E}\cdot\vec{B}$ term in Eqn. (\ref{eq:contin}) leads to a pump chemical potential on the order of 9meV and thus the impurity scattering rate should be kept below a few meV.  This impurity scattering rate is both realistic and achievable. The effect of finite temperature had a more dramatic effect on the interband transitions, so it seems likely that low temperatures will be required to resolve the steps cleanly. Optical experiments are routinely carried out at a few Kelvin and so we expect this should not pose a technical challenge. Finally, we discussed how the application of the $\vec{E}\cdot\vec{B}$ field can be used to trace out the frequency dependence of the scattering rate.  Since the $\vec{E}\cdot\vec{B}$ field controls the chemical potential, it can be used to probe $\Gamma(\omega)$ at many different energy scales.

It is natural to wonder about the effects that the gapless surface states of Weyl semimetals (known as Fermi arcs) would have on the measurement of the optical conductivity. Since the optical conductivity is a bulk probe, we expect our results to be largely unchanged by the presence of the Fermi arc surface states. The features in the interband occur at energies too high to be affected by the low energy surface states. The Fermi arcs may enhance the low energy absorption (i.e. the Drude), however, analysis of the changes to the low frequency absorption would require the study of the Weyl semimetal in the presence of a boundary. This boundary value problem is beyond the scope of the results presented here.

We would like to address one final point. In this work we considered the case of a weak magnetic field where Landau level quantization was unimportant. If one applies a $\vec{B}$ field strong enough that Landau level formation is important, the interband lineshapes discussed here should be replaced instead by the magneto-optical conductivity lineshapes\cite{Ashby:2013ys}.  The step-like features will still appear in the interband, but occur from a superposition of the magneto-optical lineshapes instead of the free Fermion lineshapes discussed here.

\begin{acknowledgements}
This work was supported by the Natural Sciences and Engineering Research Council of Canada and the Canadian Institute for Advanced Research.
\end{acknowledgements}

\appendix

\section{Calculation of Optical Conductivity}
\label{sec:apa}
\begin{widetext}
The Kubo formula reads
\begin{align}
\textrm{Re}\sigma_{ij}(\Omega) = \frac{e^2\pi}{\Omega}\int_{-\infty}^\infty d\omega [f(\omega-\mu) - f(\omega+\Omega)]\int \frac{d^3k}{(2\pi)^3} \mathrm{Tr}\left[v_i\hat{A}(\vec{k},\omega)v_j\hat{A}(\vec{k},\omega+\Omega)\right]
\end{align}
The velocity operators are given by $\vec{v} = \pd{\mathcal{H}}{\vec{k}} = v_F\vec{\sigma}$.  We can find the spectral densities from the Greens function. The inverse Greens function is given by
\begin{align}
\mathcal{G}^{-1}(z) = \left(\begin{matrix}
    z-k_z & -k_x+ik_y  \\
    -k_x-ik_y & z+k_z \\
\end{matrix}\right)
\end{align}
Inverting gives
\begin{align}
\mathcal{G}(z) = \frac{1}{|k|^2-z^2}\left(\begin{matrix}
    -k_z-z & -k_x+ik_y  \\
    -k_x-ik_y & k_z-z \\
\end{matrix}\right)
\end{align}
The spectral densities follow from the relation
\begin{align}
\mathcal{G}(z) = \int_{-\infty}^\infty d\omega \frac{\hat{A}(\omega)}{z-\omega}
\end{align}
We can read off the components of $\hat{A}$ directly.  For example,
\begin{align}
\mathcal{G}_{11} = \frac{-k_z-z}{|k|^2-z^2} = \frac{-k_z-|k|}{2|k|(|k|-z)}+\frac{-k_z+|k|}{2|k|(|k|+z)}.
\end{align}
Now defining 
\begin{align}
u^2 = \frac{1}{2}\left(1+\frac{k_z}{|k|}\right)\\
v^2 = \frac{1}{2}\left(1-\frac{k_z}{|k|}\right),
\end{align}
we have that $A_{11}  = u^2\delta(\omega-|k|)+v^2\delta(\omega+|k|)$.  Similarly, $A_{22}  = v^2\delta(\omega-|k|)+u^2\delta(\omega+|k|)$.  Now for $\sigma_{xx}$ the trace takes the form $\mathrm{Tr}\left[\sigma_x\hat{A}\sigma_x\hat{A}'\right] = A_{12}A_{12}'+A_{21}A_{21}'+A_{22}A_{11}'+A_{11}A_{22}'$.  The terms proportional to $A_{12}$ and $A_{21}$ will vanish once the angular integration is carried out since they are proportional to $k_x\pm ik_y$.  So we have $\mathrm{Tr}\left[\sigma_x\hat{A}\sigma_x\hat{A}'\right] = A_{22}A_{11}'+A_{11}A_{22}'$.  Now we define $A_{\pm} = \delta(\omega\mp\epsilon_k)$ and $A_{\pm}' = \delta(\omega+\Omega\mp\epsilon_k)$.  Expanding out the trace we have $\mathrm{Tr}\left[\sigma_x\hat{A}\sigma_x\hat{A}'\right] = 2u^2v^2\left(A_+A_+'+A_-A_-'\right)+\left(u^4+v^4\right)\left(A_-A_+'+A_+A_-'\right)$.  We will now perform the angular integration.  We need $u^4+v^4 = \frac{1}{2}\left(1+\frac{k_z^2}{|k|^2}\right)$ and $2u^2v^2 = \frac{1}{2}\left(1-\frac{k_z^2}{|k|^2}\right)$.  The relevant integrals are thus
\begin{align}
&\int\frac{d^3k}{(2\pi)^3}\frac{1}{2}\left(1\pm\frac{k_z^2}{|k|^2}\right)\\
&=\frac{1}{2}\int d\epsilon\frac{\epsilon^2}{2\pi^2v_F^3} \pm \frac{1}{2}\int d\epsilon\frac{\epsilon^2}{(2\pi)^3v_F^3}2\pi\int_0^\pi d\theta \sin(\theta)\cos^2(\theta)\\
& = \frac{3\pm1}{6v_F^3}\int d\epsilon\frac{\epsilon^2}{2\pi^2}
\end{align}
Finally we obtain 
\begin{align}
\textrm{Re}\sigma_{xx}(\Omega) = \frac{e^2}{6\pi v_F}\int_{-\infty}^\infty d\omega\frac{f(\omega-\mu)-f(\omega+\Omega-\mu)}{\Omega}\int_0^\infty d\epsilon \epsilon^2\left[A_+A_+'+A_-A_-'+2\left(A_+A_-'+A_-A_+'\right)\right].
\end{align}
In the presence of impurities, after performing an impurity average over a random distribution which restores translation invariance, the spectral functions can be written as\cite{Mahan}
\begin{align}
  A(\pm\epsilon,\omega) = \frac{1}{\pi}\frac{-\textrm{Im}\Sigma(\omega)}{\left(\omega-\textrm{Re}\Sigma(\omega)\mp\epsilon\right)^2+\left(\textrm{Im}\Sigma(\omega)\right)^2},
\end{align}
where $\Sigma(\omega)$ is the self energy. It is this self energy which in general depends on $\omega$, and carries the information on the detailed properties of the impurity potential associated with a single scattering center. It also differs depending on the strength of the impurity scattering: if the scattering is weak the self energy can be treated in the Born approximation, but if it is strong a full T-matrix approach is required, as in the unitary limit. In this paper we consider only two cases. The first is a constant residual scattering rate.  This is appropriate for a weak delta function potential with constant density of states.  The second we consider is the case considered by Burkov and Balents\cite{Burkov:2011kx} of Born scattering. In this case the scattering rate depends on the density of states and is quadratic in energy.

Using the above form for the spectral functions one can recover our formulae that contain the impurity scattering rate presented in the main text.

Now we check the clean limit at $T = 0$.  In that limit we have the intraband piece
\begin{align}
\textrm{Re}\sigma_{xx}^{\textrm{D}}(\Omega) =& \frac{e^2}{6\pi v_F\Omega}\int_{\mu-\Omega}^\mu d\omega\int_0^\infty d\epsilon \epsilon^2\left[\delta(\omega-\epsilon)\delta(\omega+\Omega-\epsilon)+\delta(\omega+\epsilon)\delta(\omega+\Omega+\epsilon)\right]\\
\label{eq:a14}& = \frac{e^2\mu^2}{6\pi v_F}\delta(\Omega) = \frac{e^2\mu^2}{3 h \hbar v_F}\delta(\Omega). 
\end{align}
We also have the interband piece
\begin{align}
\textrm{Re}\sigma_{xx}^{\textrm{IB}}(\Omega) =& \frac{e^2}{3\pi v_F\Omega}\int_{\mu-\Omega}^\mu d\omega\int_0^\infty d\epsilon \epsilon^2\left[\delta(\omega-\epsilon)\delta(\omega+\Omega+\epsilon)+\delta(\omega+\epsilon)\delta(\omega+\Omega-\epsilon)\right]\\
\label{eq:a16}& = \frac{e^2}{24\pi v_F}\Omega\Theta(\Omega-2\mu)= \frac{e^2}{12h \hbar v_F}\Omega\Theta(\Omega-2\mu).
\end{align}
These two results give the well known formula Eq. (\ref{eq:clean}).  In the final equalities we have restored the factors of $\hbar$ to make the physical units clear.
\end{widetext}
\bibliography{biblio}

\begin{thebibliography}{35}
\expandafter\ifx\csname natexlab\endcsname\relax\def\natexlab#1{#1}\fi
\expandafter\ifx\csname bibnamefont\endcsname\relax
  \def\bibnamefont#1{#1}\fi
\expandafter\ifx\csname bibfnamefont\endcsname\relax
  \def\bibfnamefont#1{#1}\fi
\expandafter\ifx\csname citenamefont\endcsname\relax
  \def\citenamefont#1{#1}\fi
\expandafter\ifx\csname url\endcsname\relax
  \def\url#1{\texttt{#1}}\fi
\expandafter\ifx\csname urlprefix\endcsname\relax\def\urlprefix{URL }\fi
\providecommand{\bibinfo}[2]{#2}
\providecommand{\eprint}[2][]{\url{#2}}

\bibitem[{\citenamefont{Wan et~al.}(2011)\citenamefont{Wan, Turner, Vishwanath,
  and Savrasov}}]{Wan:2011fk}
\bibinfo{author}{\bibfnamefont{X.}~\bibnamefont{Wan}},
  \bibinfo{author}{\bibfnamefont{A.~M.} \bibnamefont{Turner}},
  \bibinfo{author}{\bibfnamefont{A.}~\bibnamefont{Vishwanath}},
  \bibnamefont{and} \bibinfo{author}{\bibfnamefont{S.~Y.}
  \bibnamefont{Savrasov}}, \bibinfo{journal}{Physical Review B}
  \textbf{\bibinfo{volume}{83}}, \bibinfo{pages}{205101}
  (\bibinfo{year}{2011}).

\bibitem[{\citenamefont{Hosur et~al.}(2012)\citenamefont{Hosur, Parameswaran,
  and Vishwanath}}]{Hosur:2012fk}
\bibinfo{author}{\bibfnamefont{P.}~\bibnamefont{Hosur}},
  \bibinfo{author}{\bibfnamefont{S.~A.} \bibnamefont{Parameswaran}},
  \bibnamefont{and}
  \bibinfo{author}{\bibfnamefont{A.}~\bibnamefont{Vishwanath}},
  \bibinfo{journal}{Physical Review Letters} \textbf{\bibinfo{volume}{108}},
  \bibinfo{pages}{046602} (\bibinfo{year}{2012}).

\bibitem[{\citenamefont{Witczak-Krempa and Kim}(2012)}]{Witczak-Krempa:2012ve}
\bibinfo{author}{\bibfnamefont{W.}~\bibnamefont{Witczak-Krempa}}
  \bibnamefont{and} \bibinfo{author}{\bibfnamefont{Y.~B.} \bibnamefont{Kim}},
  \bibinfo{journal}{Physical Review B} \textbf{\bibinfo{volume}{85}},
  \bibinfo{pages}{045124} (\bibinfo{year}{2012}).

\bibitem[{\citenamefont{Vafek and Vishwanath}(2014)}]{vafek:2013fk}
\bibinfo{author}{\bibfnamefont{O.}~\bibnamefont{Vafek}} \bibnamefont{and}
  \bibinfo{author}{\bibfnamefont{A.}~\bibnamefont{Vishwanath}},
  \bibinfo{journal}{Annual Review of Condensed Matter Physics}
  \textbf{\bibinfo{volume}{5}}, \bibinfo{pages}{83} (\bibinfo{year}{2014}).

\bibitem[{\citenamefont{Nielsen and Ninomiya}(1981)}]{Nielsen:1981fv}
\bibinfo{author}{\bibfnamefont{H.~B.} \bibnamefont{Nielsen}} \bibnamefont{and}
  \bibinfo{author}{\bibfnamefont{M.}~\bibnamefont{Ninomiya}},
  \bibinfo{journal}{Physics Letters B} \textbf{\bibinfo{volume}{105}},
  \bibinfo{pages}{219} (\bibinfo{year}{1981}).

\bibitem[{\citenamefont{Burkov and Balents}(2011)}]{Burkov:2011kx}
\bibinfo{author}{\bibfnamefont{A.~A.} \bibnamefont{Burkov}} \bibnamefont{and}
  \bibinfo{author}{\bibfnamefont{L.}~\bibnamefont{Balents}},
  \bibinfo{journal}{Physical Review Letters} \textbf{\bibinfo{volume}{107}},
  \bibinfo{pages}{127205} (\bibinfo{year}{2011}).

\bibitem[{\citenamefont{Burkov et~al.}(2011)\citenamefont{Burkov, Hook, and
  Balents}}]{Burkov:2011ys}
\bibinfo{author}{\bibfnamefont{A.~A.} \bibnamefont{Burkov}},
  \bibinfo{author}{\bibfnamefont{M.~D.} \bibnamefont{Hook}}, \bibnamefont{and}
  \bibinfo{author}{\bibfnamefont{L.}~\bibnamefont{Balents}},
  \bibinfo{journal}{Physical Review B} \textbf{\bibinfo{volume}{84}},
  \bibinfo{pages}{235126} (\bibinfo{year}{2011}).

\bibitem[{\citenamefont{Zyuzin et~al.}(2012)\citenamefont{Zyuzin, Wu, and
  Burkov}}]{Zyuzin:2012zr}
\bibinfo{author}{\bibfnamefont{A.~A.} \bibnamefont{Zyuzin}},
  \bibinfo{author}{\bibfnamefont{S.}~\bibnamefont{Wu}}, \bibnamefont{and}
  \bibinfo{author}{\bibfnamefont{A.~A.} \bibnamefont{Burkov}},
  \bibinfo{journal}{Physical Review B} \textbf{\bibinfo{volume}{85}},
  \bibinfo{pages}{165110} (\bibinfo{year}{2012}).

\bibitem[{\citenamefont{Hal{\'a}sz and Balents}(2012)}]{Halasz:2012ly}
\bibinfo{author}{\bibfnamefont{G.~B.} \bibnamefont{Hal{\'a}sz}}
  \bibnamefont{and} \bibinfo{author}{\bibfnamefont{L.}~\bibnamefont{Balents}},
  \bibinfo{journal}{Physical Review B} \textbf{\bibinfo{volume}{85}},
  \bibinfo{pages}{035103} (\bibinfo{year}{2012}).

\bibitem[{\citenamefont{Timusk et~al.}(2013)\citenamefont{Timusk, Carbotte,
  Homes, Basov, and Sharapov}}]{Timusk:2013fk}
\bibinfo{author}{\bibfnamefont{T.}~\bibnamefont{Timusk}},
  \bibinfo{author}{\bibfnamefont{J.~P.} \bibnamefont{Carbotte}},
  \bibinfo{author}{\bibfnamefont{C.~C.} \bibnamefont{Homes}},
  \bibinfo{author}{\bibfnamefont{D.~N.} \bibnamefont{Basov}}, \bibnamefont{and}
  \bibinfo{author}{\bibfnamefont{S.~G.} \bibnamefont{Sharapov}},
  \bibinfo{journal}{Physical Review B} \textbf{\bibinfo{volume}{87}},
  \bibinfo{pages}{235121} (\bibinfo{year}{2013}).

\bibitem[{\citenamefont{Neupane et~al.}(2013)\citenamefont{Neupane, Xu, Sankar,
  Alidoust, Bian, Liu, Belopolski, Chang, Jeng, Lin et~al.}}]{Neupane:fk2013}
\bibinfo{author}{\bibfnamefont{M.}~\bibnamefont{Neupane}},
  \bibinfo{author}{\bibfnamefont{S.}~\bibnamefont{Xu}},
  \bibinfo{author}{\bibfnamefont{R.}~\bibnamefont{Sankar}},
  \bibinfo{author}{\bibfnamefont{N.}~\bibnamefont{Alidoust}},
  \bibinfo{author}{\bibfnamefont{G.}~\bibnamefont{Bian}},
  \bibinfo{author}{\bibfnamefont{C.}~\bibnamefont{Liu}},
  \bibinfo{author}{\bibfnamefont{I.}~\bibnamefont{Belopolski}},
  \bibinfo{author}{\bibfnamefont{T.-R.} \bibnamefont{Chang}},
  \bibinfo{author}{\bibfnamefont{H.-T.} \bibnamefont{Jeng}},
  \bibinfo{author}{\bibfnamefont{H.}~\bibnamefont{Lin}}, \bibnamefont{et~al.},
  \bibinfo{journal}{arXiv:1309.7892 [cond-mat.mes-hall]}
  (\bibinfo{year}{2013}).

\bibitem[{\citenamefont{Borisenko et~al.}(2013)\citenamefont{Borisenko, Gibson,
  Evtushinsky, Zabolotnyy, Buechner, and Cava}}]{Borisenko:fk2013}
\bibinfo{author}{\bibfnamefont{S.}~\bibnamefont{Borisenko}},
  \bibinfo{author}{\bibfnamefont{Q.}~\bibnamefont{Gibson}},
  \bibinfo{author}{\bibfnamefont{D.}~\bibnamefont{Evtushinsky}},
  \bibinfo{author}{\bibfnamefont{V.}~\bibnamefont{Zabolotnyy}},
  \bibinfo{author}{\bibfnamefont{B.}~\bibnamefont{Buechner}}, \bibnamefont{and}
  \bibinfo{author}{\bibfnamefont{R.~J.} \bibnamefont{Cava}},
  \bibinfo{journal}{arXiv:1309.7978 [cond-mat.mes-hall]}
  (\bibinfo{year}{2013}).

\bibitem[{\citenamefont{Xu and et~al.}(2013)}]{xu:2013nb}
\bibinfo{author}{\bibfnamefont{S.-Y.} \bibnamefont{Xu}} \bibnamefont{and}
  \bibinfo{author}{\bibfnamefont{C.~L.} \bibnamefont{et~al.}},
  \bibinfo{journal}{arXiv:1312.7624 [cond-mat.mes-hall]}
  (\bibinfo{year}{2013}).

\bibitem[{\citenamefont{Zyuzin and Burkov}(2012)}]{Zyuzin:2012fkk}
\bibinfo{author}{\bibfnamefont{A.~A.} \bibnamefont{Zyuzin}} \bibnamefont{and}
  \bibinfo{author}{\bibfnamefont{A.~A.} \bibnamefont{Burkov}},
  \bibinfo{journal}{Phys. Rev. B} \textbf{\bibinfo{volume}{86}},
  \bibinfo{pages}{115133} (\bibinfo{year}{2012}).

\bibitem[{\citenamefont{Grushin}(2012)}]{Grushin:2012fk}
\bibinfo{author}{\bibfnamefont{A.~G.} \bibnamefont{Grushin}},
  \bibinfo{journal}{Phys. Rev. D} \textbf{\bibinfo{volume}{86}},
  \bibinfo{pages}{045001} (\bibinfo{year}{2012}).

\bibitem[{\citenamefont{Aji}(2012)}]{Aji:2012fk}
\bibinfo{author}{\bibfnamefont{V.}~\bibnamefont{Aji}}, \bibinfo{journal}{Phys.
  Rev. B} \textbf{\bibinfo{volume}{85}}, \bibinfo{pages}{241101}
  (\bibinfo{year}{2012}).

\bibitem[{\citenamefont{Kim et~al.}(2013)\citenamefont{Kim, Kim, Wang, Sasaki,
  Satoh, Ohnishi, Kitaura, Yang, and Li}}]{Heon:2013fk}
\bibinfo{author}{\bibfnamefont{H.-J.} \bibnamefont{Kim}},
  \bibinfo{author}{\bibfnamefont{K.-S.} \bibnamefont{Kim}},
  \bibinfo{author}{\bibfnamefont{J.-F.} \bibnamefont{Wang}},
  \bibinfo{author}{\bibfnamefont{M.}~\bibnamefont{Sasaki}},
  \bibinfo{author}{\bibfnamefont{N.}~\bibnamefont{Satoh}},
  \bibinfo{author}{\bibfnamefont{A.}~\bibnamefont{Ohnishi}},
  \bibinfo{author}{\bibfnamefont{M.}~\bibnamefont{Kitaura}},
  \bibinfo{author}{\bibfnamefont{M.}~\bibnamefont{Yang}}, \bibnamefont{and}
  \bibinfo{author}{\bibfnamefont{L.}~\bibnamefont{Li}}, \bibinfo{journal}{Phys.
  Rev. Lett.} \textbf{\bibinfo{volume}{111}}, \bibinfo{pages}{246603}
  (\bibinfo{year}{2013}).

\bibitem[{\citenamefont{Liu et~al.}(2013)\citenamefont{Liu, Ye, and
  Qi}}]{Liu:2013fkk}
\bibinfo{author}{\bibfnamefont{C.-X.} \bibnamefont{Liu}},
  \bibinfo{author}{\bibfnamefont{P.}~\bibnamefont{Ye}}, \bibnamefont{and}
  \bibinfo{author}{\bibfnamefont{X.-L.} \bibnamefont{Qi}},
  \bibinfo{journal}{Phys. Rev. B} \textbf{\bibinfo{volume}{87}},
  \bibinfo{pages}{235306} (\bibinfo{year}{2013}).

\bibitem[{\citenamefont{Landsteiner}(2014)}]{Landsteiner:2014fk}
\bibinfo{author}{\bibfnamefont{K.}~\bibnamefont{Landsteiner}},
  \bibinfo{journal}{Phys. Rev. B} \textbf{\bibinfo{volume}{89}},
  \bibinfo{pages}{075124} (\bibinfo{year}{2014}).

\bibitem[{\citenamefont{Volovik}(2003)}]{Volovik}
\bibinfo{author}{\bibfnamefont{G.~E.} \bibnamefont{Volovik}},
  \emph{\bibinfo{title}{{The Universe in a Helium Droplet}}}
  (\bibinfo{publisher}{Oxford University Press}, \bibinfo{year}{2003}).

\bibitem[{\citenamefont{Hosur and Qi}(2014)}]{hosur:2014fk}
\bibinfo{author}{\bibfnamefont{P.}~\bibnamefont{Hosur}} \bibnamefont{and}
  \bibinfo{author}{\bibfnamefont{X.-L.} \bibnamefont{Qi}},
  \bibinfo{journal}{arXiv:1401.2762 [cond-mat.str-el]}  (\bibinfo{year}{2014}).

\bibitem[{\citenamefont{Gorbar et~al.}(2014)\citenamefont{Gorbar, Miransky, and
  Shovkovy}}]{Gorbar:2014fk}
\bibinfo{author}{\bibfnamefont{E.~V.} \bibnamefont{Gorbar}},
  \bibinfo{author}{\bibfnamefont{V.~A.} \bibnamefont{Miransky}},
  \bibnamefont{and} \bibinfo{author}{\bibfnamefont{I.~A.}
  \bibnamefont{Shovkovy}}, \bibinfo{journal}{Phys. Rev. B}
  \textbf{\bibinfo{volume}{89}}, \bibinfo{pages}{085126}
  (\bibinfo{year}{2014}).

\bibitem[{\citenamefont{Xu et~al.}(2011)\citenamefont{Xu, Weng, Wang, Dai, and
  Fang}}]{Xu:2011kx}
\bibinfo{author}{\bibfnamefont{G.}~\bibnamefont{Xu}},
  \bibinfo{author}{\bibfnamefont{H.}~\bibnamefont{Weng}},
  \bibinfo{author}{\bibfnamefont{Z.}~\bibnamefont{Wang}},
  \bibinfo{author}{\bibfnamefont{X.}~\bibnamefont{Dai}}, \bibnamefont{and}
  \bibinfo{author}{\bibfnamefont{Z.}~\bibnamefont{Fang}},
  \bibinfo{journal}{Physical Review Letters} \textbf{\bibinfo{volume}{107}},
  \bibinfo{pages}{186806} (\bibinfo{year}{2011}).

\bibitem[{\citenamefont{Goswami and Tewari}(2013)}]{Goswami:2012fk}
\bibinfo{author}{\bibfnamefont{P.}~\bibnamefont{Goswami}} \bibnamefont{and}
  \bibinfo{author}{\bibfnamefont{S.}~\bibnamefont{Tewari}},
  \bibinfo{journal}{Phys. Rev. B} \textbf{\bibinfo{volume}{88}},
  \bibinfo{pages}{245107} (\bibinfo{year}{2013}).

\bibitem[{\citenamefont{Jian-Hui et~al.}(2013)\citenamefont{Jian-Hui, Hua,
  Qian, and Jun-Ren}}]{Jian-Hui:2013kx}
\bibinfo{author}{\bibfnamefont{Z.}~\bibnamefont{Jian-Hui}},
  \bibinfo{author}{\bibfnamefont{J.}~\bibnamefont{Hua}},
  \bibinfo{author}{\bibfnamefont{N.}~\bibnamefont{Qian}}, \bibnamefont{and}
  \bibinfo{author}{\bibfnamefont{S.}~\bibnamefont{Jun-Ren}},
  \bibinfo{journal}{Chinese Physics Letters} \textbf{\bibinfo{volume}{30}},
  \bibinfo{pages}{027101} (\bibinfo{year}{2013}).

\bibitem[{\citenamefont{Chen et~al.}(2013)\citenamefont{Chen, Wu, and
  Burkov}}]{Chen:2013uq}
\bibinfo{author}{\bibfnamefont{Y.}~\bibnamefont{Chen}},
  \bibinfo{author}{\bibfnamefont{S.}~\bibnamefont{Wu}}, \bibnamefont{and}
  \bibinfo{author}{\bibfnamefont{A.~A.} \bibnamefont{Burkov}},
  \bibinfo{journal}{Physical Review B} \textbf{\bibinfo{volume}{88}},
  \bibinfo{pages}{125105} (\bibinfo{year}{2013}).

\bibitem[{\citenamefont{Parameswaran et~al.}(2013)\citenamefont{Parameswaran,
  Grover, Abanin, Pesin, and Vishwanath}}]{Parameswaran:fk2013}
\bibinfo{author}{\bibfnamefont{S.}~\bibnamefont{Parameswaran}},
  \bibinfo{author}{\bibfnamefont{T.}~\bibnamefont{Grover}},
  \bibinfo{author}{\bibfnamefont{D.~A.} \bibnamefont{Abanin}},
  \bibinfo{author}{\bibfnamefont{D.~A.} \bibnamefont{Pesin}}, \bibnamefont{and}
  \bibinfo{author}{\bibfnamefont{A.}~\bibnamefont{Vishwanath}},
  \bibinfo{journal}{arXiv:1306.1234v1 [cond-mat.str-el]}
  (\bibinfo{year}{2013}).

\bibitem[{\citenamefont{Goswami et~al.}(2014)\citenamefont{Goswami, Sharma, and
  Tewari}}]{Goswami:2014fk}
\bibinfo{author}{\bibfnamefont{P.}~\bibnamefont{Goswami}},
  \bibinfo{author}{\bibfnamefont{G.}~\bibnamefont{Sharma}}, \bibnamefont{and}
  \bibinfo{author}{\bibfnamefont{S.}~\bibnamefont{Tewari}},
  \bibinfo{journal}{arXiv:1404.2927 [cond-mat.str-el]}  (\bibinfo{year}{2014}).

\bibitem[{\citenamefont{Panfilov et~al.}(2014)\citenamefont{Panfilov, Burkov,
  and Pesin}}]{Pesin:2014fk}
\bibinfo{author}{\bibfnamefont{I.}~\bibnamefont{Panfilov}},
  \bibinfo{author}{\bibfnamefont{A.}~\bibnamefont{Burkov}}, \bibnamefont{and}
  \bibinfo{author}{\bibfnamefont{D.}~\bibnamefont{Pesin}},
  \bibinfo{journal}{arXiv:1404.2927 [cond-mat.str-el]}  (\bibinfo{year}{2014}).

\bibitem[{\citenamefont{Liu et~al.}(2010)\citenamefont{Liu, Qi, Zhang, Dai,
  Fang, and Zhang}}]{Liu14:PRB}
\bibinfo{author}{\bibfnamefont{C.-X.} \bibnamefont{Liu}},
  \bibinfo{author}{\bibfnamefont{X.-L.} \bibnamefont{Qi}},
  \bibinfo{author}{\bibfnamefont{H.~J.} \bibnamefont{Zhang}},
  \bibinfo{author}{\bibfnamefont{X.}~\bibnamefont{Dai}},
  \bibinfo{author}{\bibfnamefont{Z.}~\bibnamefont{Fang}}, \bibnamefont{and}
  \bibinfo{author}{\bibfnamefont{S.-C.} \bibnamefont{Zhang}},
  \bibinfo{journal}{Phys. Rev. B} \textbf{\bibinfo{volume}{82}},
  \bibinfo{pages}{045122} (\bibinfo{year}{2010}).

\bibitem[{\citenamefont{Li and Carbotte}(2013)}]{ZhouCarb:fk14}
\bibinfo{author}{\bibfnamefont{Z.}~\bibnamefont{Li}} \bibnamefont{and}
  \bibinfo{author}{\bibfnamefont{J.~P.} \bibnamefont{Carbotte}},
  \bibinfo{journal}{Phys. Rev. B} \textbf{\bibinfo{volume}{88}},
  \bibinfo{pages}{045414} (\bibinfo{year}{2013}).

\bibitem[{\citenamefont{Orlita et~al.}(2014)\citenamefont{Orlita, Basko,
  Zholudev, Teppe, Knap, Gavrilenko, Mikhailov, Dvoretskii, Neugebauer,
  Faugeras et~al.}}]{Orlita:2014fk}
\bibinfo{author}{\bibfnamefont{M.}~\bibnamefont{Orlita}},
  \bibinfo{author}{\bibfnamefont{D.~M.} \bibnamefont{Basko}},
  \bibinfo{author}{\bibfnamefont{M.~S.} \bibnamefont{Zholudev}},
  \bibinfo{author}{\bibfnamefont{F.}~\bibnamefont{Teppe}},
  \bibinfo{author}{\bibfnamefont{W.}~\bibnamefont{Knap}},
  \bibinfo{author}{\bibfnamefont{V.~I.} \bibnamefont{Gavrilenko}},
  \bibinfo{author}{\bibfnamefont{N.~N.} \bibnamefont{Mikhailov}},
  \bibinfo{author}{\bibfnamefont{S.~A.} \bibnamefont{Dvoretskii}},
  \bibinfo{author}{\bibfnamefont{P.}~\bibnamefont{Neugebauer}},
  \bibinfo{author}{\bibfnamefont{C.}~\bibnamefont{Faugeras}},
  \bibnamefont{et~al.}, \bibinfo{journal}{Nature Physics}
  \textbf{\bibinfo{volume}{10}}, \bibinfo{pages}{233} (\bibinfo{year}{2014}).

\bibitem[{\citenamefont{Basov and Timusk}(2005)}]{RevModPhys.77.721}
\bibinfo{author}{\bibfnamefont{D.~N.} \bibnamefont{Basov}} \bibnamefont{and}
  \bibinfo{author}{\bibfnamefont{T.}~\bibnamefont{Timusk}},
  \bibinfo{journal}{Rev. Mod. Phys.} \textbf{\bibinfo{volume}{77}},
  \bibinfo{pages}{721} (\bibinfo{year}{2005}).

\bibitem[{\citenamefont{Ashby and Carbotte}(2013)}]{Ashby:2013ys}
\bibinfo{author}{\bibfnamefont{P.~E.~C.} \bibnamefont{Ashby}} \bibnamefont{and}
  \bibinfo{author}{\bibfnamefont{J.~P.} \bibnamefont{Carbotte}},
  \bibinfo{journal}{Physical Review B} \textbf{\bibinfo{volume}{87}},
  \bibinfo{pages}{245131} (\bibinfo{year}{2013}).

\bibitem[{\citenamefont{Mahan}(1993)}]{Mahan}
\bibinfo{author}{\bibfnamefont{G.~D.} \bibnamefont{Mahan}},
  \emph{\bibinfo{title}{{Many-Particle Physics}}} (\bibinfo{publisher}{Plenum},
  \bibinfo{address}{New York, N.Y.}, \bibinfo{year}{1993}),
  \bibinfo{edition}{2nd} ed.

\end{thebibliography}

\end{document}